# Weight Optimization for Distributed Average Consensus Algorithm in Symmetric, CCS & KCS Star Networks

Saber Jafarizadeh, *Student Member, IEEE,* and Abbas Jamalipour, *Fellow, IEEE*

*Abstract*—This paper addresses weight optimization problem in distributed consensus averaging algorithm over networks with symmetric star topology. We have determined optimal weights and convergence rate of the network in terms of its topological parameters. In addition, two alternative topologies with more rapid convergence rates have been introduced. The new topologies are Complete-Cored Symmetric (CCS) star and *K*-Cored Symmetric (KCS) star topologies. It has been shown that the optimal weights for the edges of central part in symmetric and CCS star configurations are independent of their branches. By simulation optimality of obtained weights under quantization constraints have been verified.

## I. INTRODUCTION

IN distributed consensus problem a group of nodes (agents, sensors), have to reach a common decision in a distributed fashion. Some of its major applications include distributed agreement, synchronization problem [1], multi-vehicle control and navigation [2] and load balancing in parallel processors [3].

Distributed consensus algorithm in its most simple form reduces to distributed consensus averaging algorithm, where the nodes have to compute the average of their initial states. Distributed consensus averaging problem is an inevitable part of the solution for more complex problems in several applications. Some of these applications are multiagent distributed coordination and flocking [4, 5], distributed data fusion in sensor networks, [6], fastest mixing Markov chain problem, [7], gossip algorithms [8], sensor localization, [9] and distributed estimation and detection for decentralized sensor networks, [10, 11].

In distributed consensus averaging algorithm each node updates its state by a weighted average of its own and neighbors' states. Convergence rate of the algorithm depends on the choice of weights. One of major challenges in the field of distributed consensus averaging algorithms is the design of weighting methods. Researchers have studied this problem for two different categories of networks; networks with time-varying topology and networks with fixed topology.

In networks with time varying topology, the main challenge is to select weights that guarantee the convergence of distributed consensus averaging algorithm (see [4, 10] and references therein). While in networks with fixed topology the main problem is designing the optimal weights which result in the fastest possible convergence rate. This problem is known as Fastest Distributed Consensus (FDC) averaging problem (see [12, 13] and references therein).

In previous works, the weight optimization in FDC averaging problem was treated by numerical convex optimization methods. Therefore no closed-form solution for finding the optimal weights was proposed. The only exceptions are [14, 15, 16, 17] which have considered path, two fused star, petal and star-mesh hybrid topologies, respectively.

Here in this work we have solved FDC averaging problem analytically over a sensor network with symmetric star topology. We have provided closed formed formulas for optimal weights and the convergence rate of algorithm. The solution procedure includes stratification and semidefinite programming. Using stratification method we have been able to reduce the number of variables in the semidefinite programming formulation. While by enforcing the complementary slackness conditions in semidefinite programming we obtain the characteristic polynomials, which in turn results in optimal weights and the convergence rate of algorithm. Furthermore, we have introduced two topologies with more rapid convergence rates, as alternatives to the symmetric star topology. These new topologies are Complete Cored Symmetric (CCS) star and *K*-Cored Symmetric (KCS) star topologies. It has been shown that the optimal weight for the edges connected to central node in a symmetric star configuration is independent of type of the branches connected to central node. In other words when the path branches in symmetric star topology are replaced by any arbitrary graph, the optimal weights on the edges connected to central node remain unchanged. Same conclusion holds true about the optimal weight for edges of complete core in CCS star topology. We have simulated consensus averaging algorithm with different weighting methods under quantization constraint. These simulations have been done in an effort to have a more realistic comparison between optimal weights and other common weighting methods. The weighting methods (other than optimal weights), we have considered for our simulations are maximum degree [12], Metropolis-Hasting [7] and best constant [1] weighting methods. The obtained results confirm that the quantized distributed consensus algorithm with optimal weights converges substantially faster than the one with other weighting methods.

The organization of the paper is as follows. In section II we briefly review the literature on the agreement and consensus algorithms. In Section III we review the distributed consensus

S. Jafarizadeh and A. Jamalipour are with the School of Electrical and Information Engineering, University of Sydney, Sydney NSW 2006, Australia (e-mail: saber.jafarizadeh@sydney.edu.au, abbas.jamalipour@sydney.edu.au).

averaging algorithm. In Section IV we state our main results, including definition of symmetric star topology along with two introduced topologies, namely CCS star and KCS star topologies. Furthermore the corresponding evaluated convergence rate and optimal weights of these topologies are presented in this section. Section V is devoted to the proof of our main results. Section VI presents simulation results. In this section we have compared optimal weights with other common weighting methods, under quantization constraints. We conclude the paper with discussion in Section VII.

## II. RELATED WORK

The early works in distributed computation and consensus problem include [18, 19, 20]. Tsitsiklis in [20] provided a systematic analysis for generic agreement algorithms in an asynchronous distributed environment.

References [4, 21, 22, 23], have studied distributed consensus algorithm over sensor networks with random link failures and time-varying topology. Authors in [4], has implemented a continuous time state update model for distributed consensus algorithm with delayed communication. In [23] the authors have shown that in networks with time-varying topologies, Metropolis weights converge to the mean value provided that the network's connectivity graph remains connected during iterations.

Many works has studied distributed consensus in presence of noise (see [24, 25, 26, 27], and references therein). In [24] authors have defined a Lyapunov function to formulate the mean square convergence of distributed consensus algorithm. [25] studied consensus averaging with packet-dropping communication links. Kar and Moura [26] have considered average consensus with random topologies and noisy channels. They have proposed two different algorithms to reduce the mean and variance of error at the same time. Schizas et al [27] have shown that consensus with intention of distributed estimation over ad hoc WSNs with quantization and noisy channels can be expressed as the solution of multiple constrained convex optimization sub-problems.

Weight optimization problem over networks with fixed topology has been addressed by several researchers (see [7, 12, 13, 14, 15, 16, 17]). Boyd et al [7] have provided the necessary and sufficient conditions for convergence of distributed consensus averaging algorithm as following. The largest eigenvalue of weight matrix in magnitude should be one and the corresponding eigenvector have to be the vector with all elements equal to one. They have shown that the eigenvalue with the second largest magnitude (after one) will determine the convergence rate of FDC averaging algorithm. In [7] this parameter has been denoted as the Second Largest Eigenvalue Modulus (SLEM) of the weight matrix. Furthermore they have formulated FDC averaging problem as a convex optimization problem, in particular a semidefinite program. In [13] the symmetry of network's connectivity graph has been employed to reduce the complexity of semidefinite formulation of FDC averaging problem.

Quantized consensus has been studied by several authors (see [28, 29, 30, 31, 32] and references therein). In [28] Yildiz and Scaglione have modeled the quantization error as an additional noise and by estimating it they have reduced the variance of the quantization error. In [29] Kashyap et al. have studied the quantized consensus averaging algorithm under the constraint that average is preserved at every iteration. They have proved that the necessary condition for such a constraint is that product of the quantization resolution and the number of nodes must be a factor of the average of states. However, even when average is preserved at every iteration, all nodes in the network do not necessarily converge to same value. Carli et al. [31] propose to round each node value at each step to the nearest integer. Their method results in the same outcome as the method proposed in [29]. In [32] Aysal et al. have employed probabilistic quantization (PQ) as the quantization scheme. Using this method the convergence of nodes in assured but the average is not preserved.

## III. DISTRIBUTED CONSENSUS AVERAGING ALGORITHM

This section presents a brief review of distributed consensus averaging algorithm.

Distributed consensus averaging algorithm intends to compute the average of initial states of node, $\bar{x} = (\mathbf{1}\mathbf{1}^T/n)x(0)$, by using local communication between neighboring nodes. $x(0)$ is the vector of initial states of nodes and $\mathbf{1}$ denotes the column vector with all coefficients one. At each iteration the state of each node updates according to

$$x(t+1) = Wx(t). \quad (1)$$

$W$ is the weight matrix with the same sparsity pattern as the adjacency matrix of network's associated connectivity graph. $t = 0,1,2,...$ is the discrete time index and $n$ is the number of nodes in the network.

Linear iteration (1) converges to the mean value iff one is a simple eigenvalue of $W$ associated with the eigenvector $\mathbf{1}$, and all other eigenvalues are strictly less than one in magnitude [12]. FDC averaging problem can be formulated as the following minimization problem

$$\min_{W} \quad \max(\lambda_2, -\lambda_n)$$
$$s.t. \quad W = W^T, W\mathbf{1} = \mathbf{1}, \forall \{i,j\} \notin \mathcal{E}: W_{ij} = 0$$

$1 = \lambda_1 \geq \lambda_2 \geq \cdots \geq \lambda_n \geq -1$ are eigenvalues of $W$ arranged in decreasing order and $\max(\lambda_2, -\lambda_n)$ is the *Second Largest Eigenvalue Modulus* (*SLEM*) of $W$. The main problem can be formulated in the semidefinite programming form as [12]:

$$\min_{W} \quad s$$
$$s.t. \quad -sI \preccurlyeq W - \mathbf{1}\mathbf{1}^T/n \preccurlyeq sI, W = W^T \quad (2)$$
$$W\mathbf{1} = \mathbf{1}, \forall \{i,j\} \notin \mathcal{E}: W_{ij} = 0.$$

We refer to problem (2) as the Fastest Distributed Consensus (FDC) averaging problem.

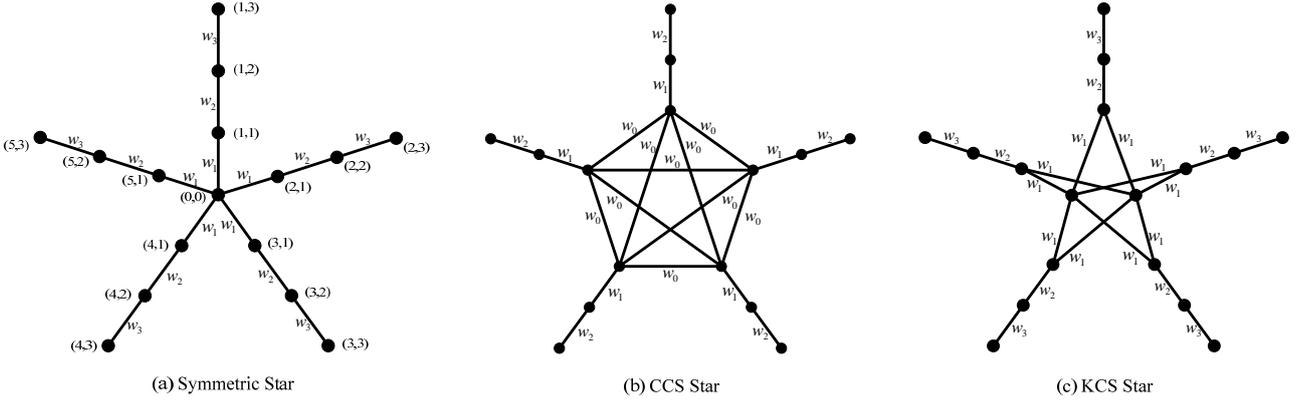

(a) Symmetric Star    (b) CCS Star    (c) KCS Star

Fig. 1. (a) Weighted symmetric star topology with $n = 5, m = 3$. (b) Weighted CCS star topology with $n = 5, m = 2$. (c) Weighted KCS star topology with $n = 5, m = 3, k = 2$.

## IV. MAIN RESULTS

This section presents the main results of the paper. Here we introduce symmetric star topology along with two alternative topologies, namely Complete Cored Symmetric (CCS) star and $K$-Cored Symmetric (KCS) star topologies. The corresponding evaluated *SLEM* and optimal weights of these topologies are presented in this section as well. Proofs and more detailed discussion are deferred to section V.

### A. Symmetric Star Topology

A sensor network with symmetric star topology of order $(m, n)$ consists of $n$ path branches of length $m$, connected to one central node. We call the path branches tails. Each one of path branches contains $m$ edges. A symmetric star graph of order $n = 5, m = 3$ is depicted in Fig. 1(a).

The optimal weights for the edges connecting tails to the central node (weighted by $w_1$ in Fig. 1(a)) equal

$$w_1 = 2/(n + 2). \tag{3}$$

For the rest of the edges the optimal weights equal,

$$w_i = 1/2. \tag{4}$$

SLEM of symmetric star topology equals $\cos(\theta)$ where $\theta$ is the smallest root of

$$(n-2)\cos\left(\left(m-\frac{1}{2}\right)\theta\right) = (n+2)\cos\left(\left(m+\frac{1}{2}\right)\theta\right), \tag{5}$$

in the interval $(0, \pi)$.

It is obvious that in symmetric star topology, the central node acts as the bottleneck of network and decreases the convergence rate of network. In the following we describe two alternative topologies with smaller SLEM values and consequently more rapid convergence rates. The topologies are Complete Cored Symmetric (CCS) star and $K$-Cored Symmetric (KCS) star topologies. Symmetric star topology can be transformed into one of these topologies (which have more rapid convergence rate) by adding a few numbers of nodes and branches to the central core of network. Therefore by adding a few numbers of nodes and branches to symmetric star topology we can enhance the convergence rate of network. This the main reason for our interest in CCS and KCS star topologies.

In CCS star topology, the central node is replaced by a complete graph, which we call it complete core. Complete core acts like a fusion core which immediately fuses (averages) all information received from branches. Therefore its SLEM is independent of number of branches. This is the main advantage of CCS star topology. Nevertheless replacing a central node by its equivalent complete core requires the addition of $n(n-3)/2$ edges. For large values of $n$, it is not practical to add $n(n-3)/2$ connections. In this case KCS topology would be a better option. In KCS topology one central node is replaced by $k$ parallel central nodes to reduce the bottleneck effect of central node. In KCS topology for adding each parallel central node only $n$ new connections are required.

### B. Complete Cored Symmetric (CCS) Star Topology

CCS star topology of order $(m, n)$ consists of $n$ path branches of length $m$, called tails. Each one of path branches contains $m$ edges. Tails are connected to each other at one end to form a complete graph in the core. A CCS star graph of order $n = 5, m = 2$ is depicted in Fig. 1(b).

The optimal weight for the edges on the complete part (weighted by $w_0$ in Fig. 1(b)) equals

$$w_0 = 1/n. \tag{6}$$

For the rest of the edges the optimal weights equal,

$$w_i = 1/2. \tag{7}$$

SLEM of CCS star topology equals

$$\text{SLEM} = \cos(\pi/(2(m+1))). \tag{8}$$

**Remark 1.** According to (8) SLEM of CCS star topology is independent of number of branches $(n)$. This means that one can add as many branches to CCS star topology without effecting SLEM of network. This comes at the cost of adding

$n$ new edges in the core for increasing the number of branches from $n$ to $n+1$.

## C. K-Cored Symmetric (KCS) Star Topology

A KCS star graph of order $(m,n,k)$ is a symmetric star graph of order $(m,n)$ with $k$ parallel central nodes instead of one. In KCS star graph there isn't any direct connection between central nodes. A KCS star graph of order $n=5, m=3$ and $k=2$ is depicted in Fig. 1(c).

The optimal weight for the edges connecting tails to $k$ parallel central nodes (weighted by $w_1$ in Fig. 1(c)) equals

$$w_1 = 2/(n+2k). \quad (9)$$

For the rest of the edges the optimal weights equal,

$$w_i = 1/2. \quad (10)$$

SLEM of KCS star topology equals $\cos(\theta)$ where $\theta$ is the smallest root of

$$(\cos(\theta)(n+2k) - n)\sin((m+1)\theta) = 2k\sin(m\theta), \quad (11)$$

in the interval $(0,\pi)$.

**Remark 2.** The optimal weights (9) are obtained by ignoring the single eigenvalue $1 - nw_1$. Therefore these results are true for the values of $k$ where $1 - nw_1$ is smaller than SLEM. This is equivalent to the following inequality

$$(2k-n)/(2k+n) \leq SLEM.$$

The smallest and largest values of $k$ satisfying the inequality above are one and $k_{max}$, respectively. There isn't any closed formed formula for $k_{max}$. The values of $k_{max}$ for different lengths $(m)$ and number $(n)$ of branches are presented in Table 1. These results are obtained numerically.

Table 1. $k_{max}$ in terms of length $(m)$ and number $(n)$ of branches

| $m$ \ $n$ | 1 | 2 | 3 | 4 | 5 | 6 | 7 | 8 |
|---|---|---|---|---|---|---|---|---|
| 2 | 2 | 7 | 15 | 26 | 41 | 58 | 79 | 104 |
| 3 | 3 | 10 | 22 | 39 | 61 | 87 | 119 | 155 |
| 4 | 4 | 13 | 29 | 52 | 81 | 116 | 158 | 207 |
| 5 | 5 | 16 | 36 | 64 | 101 | 145 | 198 | 259 |
| 6 | 6 | 20 | 43 | 77 | 121 | 174 | 237 | 310 |
| 7 | 7 | 23 | 50 | 90 | 141 | 203 | 277 | 362 |
| 8 | 8 | 26 | 58 | 103 | 161 | 232 | 316 | 413 |
| 9 | 9 | 29 | 65 | 115 | 181 | 261 | 356 | 465 |
| 10 | 10 | 32 | 72 | 128 | 201 | 290 | 395 | 517 |
| 11 | 11 | 36 | 79 | 141 | 221 | 319 | 435 | 568 |

**Remark 3.** We are not interested in values of $k$ greater than $k_{max}$. Since SLEM of KCS topology increases as $k$ gets larger than $k_{max}$. In Fig. 2 SLEM of KCS topology for $n=3, m=2$ is depicted in terms of $k$ (number of parallel central nodes). The optimal values of weights are calculated numerically. For values of $k$ greater than $k_{max}$, the optimal weights are the same except $w_1$ and $w_2$. As it is obvious from Fig. 2 for $k$ equal to $k_{max}$ we obtain the minimum possible value of SLEM for KCS topology.

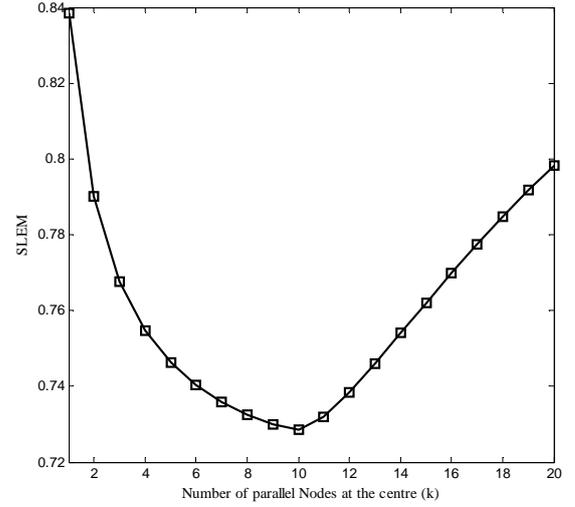

Fig. 2. SLEM of KCS topology for $n=3, m=2$ in terms of number of parallel nodes at center $(k)$.

It would be of great interest to have a comparison between the convergence rates of three topologies introduced in this section. Therefore we consider path branches of length $m=2$ (each with two edges and three nodes) and we connect them to each other in three configurations introduced in this section. The resultant topologies are symmetric star with $m=3$, CCS star with $m=2$ and KCS star with $m=3, k=2$. Their SLEM values are listed in Table 2 for different number of branches.

Table 2. SLEM of symmetric star with $m=3$, CCS star with $m=2$ and KCS star with $m=3, k=2$ for different number of branches.

| Number of Branches $(n)$ | Symmetric star | CCS star | KCS star |
|---|---|---|---|
| 3 | 0.91294 | 0.866025 | 0.893816 |
| 40 | 0.984946 | 0.866025 | 0.972613 |

From Table 2 we can see that by adding a parallel central node to network (KCS configuration), the convergence rate of network increases, but it doesn't reach the convergence rate of CCS configuration. In fact CCS configuration has smaller SLEM value than its equivalent KCS topology regardless of number of parallel nodes in the center.

## D. Symmetric, CCS & KCS Star Topologies with Path-Like branches

In [33] four branches other than path branch are introduced with their corresponding optimal weights. These branches are Lollipop, Semi-Complete, Ladder and Palm branches. It has been proved that the optimal weights presented in [33] are independent of the rest of network. These branches can be used in all three configurations described previously in this section, while their optimal weights would remain the same. In the following we have provided an example of such topologies.

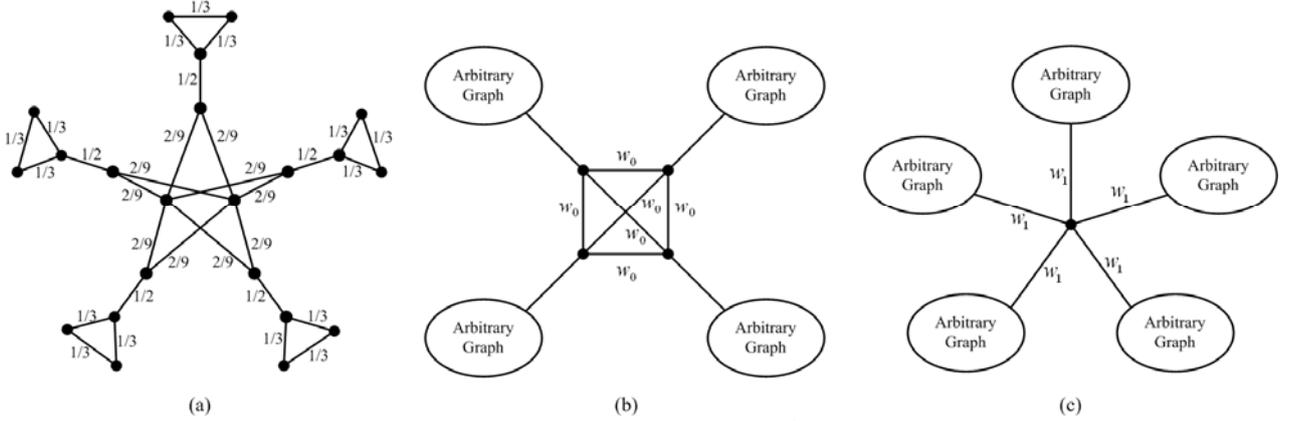
Fig. 3. (a) KCS star graph with two parallel central nodes and five Lollipop branches of order (2,3). (b) Generic CCS star Topology. (c) Generic symmetric star topology.

*Example.* In Fig. 3(a) a KCS star topology with two parallel central nodes and five Lollipop branches of order (2,3) is depicted. The optimal weights are as illustrated in Fig. 3(a). SLEM of this topology is 0.9374.

### E. Optimal weights of central edges in generic CCS & symmetric star network

In generic CCS and symmetric star configurations path branches can be replaced with any other arbitrary graph. In such configurations the optimal weights on edges of the central part would be the same as in CCS and symmetric star topology with path branches. In other words, the optimal weight for the edges of complete core (weighted as $w_0$ in Fig.3(b)) in generic CCS star topology is equal to

$$w_0 = \frac{1}{\text{Number of Branches}},$$

and the optimal weight for the edges connected to central node (weighted as $w_1$ in Fig. 3(c)) in generic symmetric star configuration is

$$w_1 = \frac{2}{2 + \text{Number of Branches}}.$$

## V. PROOF OF MAIN RESULTS

In this section we provide the solution of FDC averaging problem for sensor networks with symmetric star topology (introduced in section IV). Due to the lack of space, we omit the proof for CCS and KCS star topologies, rather we present only the proof for symmetric star topology.

We model the connectivity of sensor network with the undirected graph $G = (\mathcal{V}, \mathcal{E})$. $\mathcal{V}$ and $\mathcal{E}$ are the sets of nodes and edges, respectively. This graph $(G)$ is called the associated connectivity graph of network.

*Symmetric Star Topology*

A symmetric star graph has $|\mathcal{V}| = 1 + nm$ nodes and $|\mathcal{E}| = nm$ edges (see Fig. 1(a) for $n = 5$, $m = 3$). We denote the nodes of the graph by $\mathcal{V} = \{(i,j)|i = 1,...,n, j = 1,...,m\} \cup \{(0,0)\}$.

The automorphism group of symmetric star graph $Aut(G)$ is isomorphic to $S_n$ permutation of tails. Symmetric star graph has $(m+1)$ orbits acting on vertices and $m$ edge orbits. The orbits of $Aut(G)$ acting on the vertices are

$$\{(0,0)\} \cup \{(1,j),(2,j),...,(n,j)| j = 1,...,m\},$$

and the edge orbits of $Aut(G)$ are

$$\{((0,0),(i,1))|i = 1,...,n\},$$

and

$$\{((i,j-1),(i,j))|i = 1,...,n,\} \text{ for } j = 2,...,m.$$

*Stratification of Symmetric Star Topology*

We refer the reader to [15] (section III) for a review of stratification method and derivation of semidefinite programming.

The weights on the edges within an orbit or stratum are the same [15]. Therefore it suffices to consider just $m$ weights $w_1, w_2, ..., w_m$ (as labeled in Fig. 1(a). for $m = 3$). We associate with the node $(i,j)$, the $|\mathcal{V}| \times 1$ column vector $e_{i,j}$ defined as

$$e_{i,j} = e_i \otimes e_j \text{ for } \{i,j\} = \{i = 1,...,n, j = 1,...,m\}. \quad (12)$$

$e_i$ and $e_j$ are $n \times 1$ and $m \times 1$ column vectors with one in the $i$-th and $j$-th position respectively and zero elsewhere. We denote the central node $(0,0)$ by $e_{0,0}$ with one in the last position and zeros elsewhere. The vectors $e_{i,j}$ defined in (12) form an orthonormal basis of $\mathbf{R}^{nm}$. Using orthonormal basis (12) the weight matrix can be written as

$$W_{(i,j),(\mu,\rho)} = \begin{cases} w_1 & \text{if } i = j = 0, \ \mu = 1,...,n, \ \rho = 1, \\ w_1 & \text{if } i = 1,...,n, j = 1, \ \mu = \rho = 0, \\ w_\rho & \text{if } i = \mu = 1,...,n, \ j + 1 = \rho = 2,...,m, \\ w_j & \text{if } i = \mu = 1,...,n, \ j = \rho + 1 = 1,...,m, \\ 1 - nw_1 & \text{if } i = j = \mu = \rho = 0, \\ 1 - w_j - w_{j+1} & \text{if } i = \mu = 1,...,n, \ j = \rho = 1,...,m-1, \\ 1 - w_m & \text{if } i = \mu = 1,...,n, \ j = \rho = m, \\ 0 & Otherwise. \end{cases} \quad (13)$$

The purpose of stratification method is to convert the weight matrix into a form which is easier to analyze and determine its SLEM. The conversion of weight matrix is achievable through changing the basis which the weight matrix is defined. To this aim we use the unitary DFT matrix of size $n \times n$ in each edge stratum to map (12) into a new set of orthonormal vectors. The new set of orthonormal basis are defined as

$$\varphi_{i,\mu} = \frac{1}{\sqrt{n}} \sum_{p=1}^{n} \omega^{\mu(p-1)} e_{p,i} \quad \text{for} \quad \begin{matrix} i = 1, \dots, m, \\ \mu = 0, \dots, n-1, \end{matrix} \quad (14)$$

with $\omega = exp(j\, 2\pi/n)$ and $\varphi_0 = e_{0,0}$. The weight matrix in the new basis takes the following block diagonal form

$$W = diag(W_0, W_1, \dots, W_{n-1}).$$

Matrices $W_i$ for $i = 0, \dots, n-1$ are as follows.

$W_0 =$
$$\begin{bmatrix} 1 - nw_1 & \sqrt{n}w_1 & 0 & \cdots & 0 \\ \sqrt{n}w_1 & 1 - w_1 - w_2 & w_2 & & \vdots \\ 0 & w_2 & 1 - w_2 - w_3 & \ddots & 0 \\ \vdots & & \ddots & \ddots & w_m \\ 0 & \cdots & 0 & w_m & 1 - w_m \end{bmatrix},$$
(15-a)

$W_1 = \cdots = W_{n-1} =$
$$\begin{bmatrix} 1 - w_1 - w_2 & w_2 & 0 & \cdots & 0 \\ w_2 & 1 - w_2 - w_3 & w_3 & & \vdots \\ 0 & w_3 & 1 - w_3 - w_4 & \ddots & 0 \\ \vdots & & \ddots & \ddots & w_m \\ 0 & \cdots & 0 & w_m & 1 - w_m \end{bmatrix}.$$
(15-b)

Using the *Cauchy Interlacing* theorem (provided in appendix A), and by taking into consideration that $W_1$ is a submatrix of $W_0$, we can state the following corollary for the eigenvalues of $W_0$ and $W_1$.

*Corollary*
If we consider $W_0$ and $W_1$ as in (15), then *Cauchy interlacing theorem* (Appendix A) implies the following relations between the eigenvalues of $W_0$ and $W_1$,

$$\lambda_{m+1}(W_0) \leq \lambda_m(W_1) \leq \cdots \leq \lambda_2(W_1) \leq \lambda_2(W_0) \leq \lambda_1(W_1) \leq \lambda_1(W_0) = 1$$

It is obvious from above relations that second largest eigenvalue $(\lambda_2(W))$ of weight matrix is the largest eigenvalue of $W_1$, while smallest eigenvalue $(\lambda_{1+nm}(W))$ of weight matrix is the smallest eigenvalue of $W_1$.

*Determination of Optimal Weights via Semidefinite Programming*
Based on the corollary above, and following section III, FDC averaging problem for a sensor network with symmetric star topology can be expressed in the form of semidefinite programming as follows.

$$\begin{aligned} min \quad & s \\ s.t. \quad & W_1 \leq sI_m, \\ & -sI_{m+1} \leq W_0. \end{aligned} \quad (16)$$

$W_0$ and $W_1$ can be written as a linear combination of rank one matrices as follows.

$$W_0 = I_{m+1} - \sum_{i=1}^{m} w_i \boldsymbol{\beta}_i \boldsymbol{\beta}_i^T, \quad (17\text{-a})$$

$$W_1 = I_m - \sum_{i=1}^{m} w_i \boldsymbol{\alpha}_i \boldsymbol{\alpha}_i^T. \quad (17\text{-b})$$

The vectors $\boldsymbol{\alpha}_i$ and $\boldsymbol{\beta}_i$ are $m \times 1$ and $(m+1) \times 1$ column vectors, respectively. They are provided in Appendix B.

Using decompositions (17), the constraints in (16) can be written as

$$sI_m - I_m + \sum_{i=1}^{m} w_i \boldsymbol{\alpha}_i \boldsymbol{\alpha}_i^T \geq 0 \quad (18\text{-a})$$

$$sI_{m+1} + I_{m+1} - \sum_{i=1}^{m} w_i \boldsymbol{\beta}_i \boldsymbol{\beta}_i^T \geq 0 \quad (18\text{-b})$$

In the following we formulate problem (16) in the form of standard semidefinite programming (described in [15]). Problem parameters $(F_i, c)$ are defined as

$$F_0 = \begin{bmatrix} -I_m & 0 \\ 0 & I_{m+1} \end{bmatrix}, \quad F_{m+1} = I_{2m+1},$$

$$F_i = \begin{bmatrix} \boldsymbol{\alpha}_i \boldsymbol{\alpha}_i^T & 0 \\ 0 & -\boldsymbol{\beta}_i \boldsymbol{\beta}_i^T \end{bmatrix} \quad \text{for} \quad i = 1, \dots, m,$$

$$c_i = 0 \quad \text{for} \quad i = 1, \dots, m, \quad c_{m+1} = 1.$$

Minimization variable $(x)$ is defined as

$$x^T = [w_1, w_2, \dots, w_m, s].$$

In the case of dual problem we choose the dual variable $Z$ as $Z = \begin{bmatrix} z_1 \\ z_2 \end{bmatrix} [z_1^T \ z_2^T]$ to ensure that $Z$ is positive definite. $z_1$ and $z_2$ are column vectors, with $m$ and $m+1$ elements, respectively. From the constraints of dual problem we obtain:

$$(\boldsymbol{\alpha}_i^T z_1)^2 = (\boldsymbol{\beta}_i^T z_2)^2 \quad \text{for} \quad i = 1, \dots, m. \quad (19)$$

Considering the optimal values of primal feasible point $(x)$ and dual feasible point $(Z)$, from complementary slackness condition $(F(x)Z = ZF(x) = 0)$, we can conclude the following relations.

$$(sI_m - W_1)z_1 = 0, \quad (sI_{m+1} + W_0)z_2 = 0 \quad (20)$$

The vectors $\boldsymbol{\alpha}_i$ and $\boldsymbol{\beta}_i$ for $i = 1, ..., m$ (provided in Appendix B) form a basis for vector spaces $R^m$ and $R^{m+1}$, respectively. Therefore we can expand $z_1$ and $z_2$ in terms of $\boldsymbol{\alpha}_i$ and $\boldsymbol{\beta}_i$ as following

$$z_1 = \sum_{i=1}^{m} a_i \boldsymbol{\alpha}_i, \qquad z_2 = \sum_{i=1}^{m} b_i \boldsymbol{\beta}_i, \qquad (21)$$

with the coordinates $a_i$ and $b_i$ for $i = 1, ..., m$ to be determined.

Substituting expansions (17) and (21) in slackness conditions (20) and comparing the coefficients of $\boldsymbol{\alpha}_i$ and $\boldsymbol{\beta}_i$ we have

$$(-s+1)a_i = w_i \boldsymbol{\alpha}_i^T z_1, \qquad (s+1)b_i = w_i \boldsymbol{\beta}_i^T z_2. \qquad (22)$$

Equation (22) holds for $i = 1, ..., m$. From (22) and dual constraints (19), we can deduce that $(-s+1)^2 a_i^2 = (s+1)^2 b_i^2$ for $i = 1, ..., m$, which is equivalent to

$$(a_i/a_j)^2 = (b_i/b_j)^2 : \forall i, j = 1, ..., m. \qquad (23)$$

We can determine $\boldsymbol{\alpha}_i^T z_1$ and $\boldsymbol{\beta}_i^T z_2$ in terms of coefficients $a_i$ and $b_i$ as follows:

$$\boldsymbol{\alpha}_i^T z_1 = \sum_{j=1}^{m} a_j G_{i,j}, \qquad \boldsymbol{\alpha}_i^T z_2 = \sum_{j=1}^{m} b_j G'_{i,j}, \qquad (24)$$

where $G$ and $G'$ are Gram matrices, defined as $G_{i,j} = \boldsymbol{\alpha}_i \boldsymbol{\alpha}_j^T$ and $G'_{i,j} = \boldsymbol{\beta}_i \boldsymbol{\beta}_j^T$, respectively. By substituting expansions (24) into (22) we achieve the following recursive equations,

$$(-s+1)a_1 = w_1(a_1 - a_2), \qquad (25\text{-a})$$

$$(-s+1)a_i = w_i(-a_{i-1} + 2a_i - a_{i+1}), \qquad (25\text{-b})$$

$$(-s+1)a_m = w_m(-a_{m-1} + 2a_m), \qquad (25\text{-c})$$

and

$$(s+1)b_1 = w_1((n+1)b_1 - b_2), \qquad (26\text{-a})$$

$$(s+1)b_i = w_i(-b_{i-1} + 2b_i - b_{i+1}), \qquad (26\text{-b})$$

$$(s+1)b_m = w_m(-b_{m-1} + 2b_m). \qquad (26\text{-c})$$

Equations (25-b) and (26-b) hold for $i = 2, ..., m-1$.

In the following we find the optimal values of SLEM ($s$), weights ($w_i$) and the coordinates ($a_i, b_i$) in an inductive manner.

In first stage, from equations (25-c) and (26-c) one can determine $a_{m-1}$ and $b_{m-1}$ in terms of $a_m$ and $b_m$ and using relation (23) we obtain

$$(-s + 1 - 2w_m)^2 = (s + 1 - 2w_m)^2$$

which results in $w_m = 1/2$. Assuming $s = \cos(\theta)$ and substituting $w_m = 1/2$ into (25-c) and (26-c) we can conclude that

$$a_{m-1} = \frac{\sin(2\theta)}{\sin(\theta)} a_m, \qquad (27\text{-a})$$

$$b_{m-1} = \frac{\sin(2(\pi - \theta))}{\sin(\pi - \theta)} b_m. \qquad (27\text{-b})$$

In the second stage by substituting $a_{m-1}$ and $b_{m-1}$ from (27) into (25-b) and (26-b) and using relation (23) (with $i = m$ and $j = m - 2$) we obtain

$$w_{m-1} = 1/2,$$

and consequently

$$a_{m-2} = \frac{\sin(3\theta)}{\sin(\theta)} a_m, \qquad (28\text{-a})$$

$$b_{m-2} = \frac{\sin(3(\pi - \theta))}{\sin(\pi - \theta)} b_m. \qquad (28\text{-b})$$

Continuing the above procedure up to the $(m - 1)$-th stage we achieve

$$w_i = 1/2, \qquad (29)$$

and

$$a_i = \frac{\sin((m - i + 1)\theta)}{\sin(\theta)} a_m, \qquad (30\text{-a})$$

$$b_i = \frac{\sin((m - i + 1)(\pi - \theta))}{\sin(\pi - \theta)} b_m, \qquad (30\text{-b})$$

where (29) and (30) hold for $i = 1, ..., m$. In the last stage, using (25-a) and (26-a) while considering (23) we have

$$w_1 = 2/(n + 2). \qquad (31)$$

Substituting $a_1$ and $a_2$ in terms of $a_m$ (30-a) and $w_1 = 2/(n + 2)$ in (25-a), we can conclude that $\theta$ has to satisfy (5) and SLEM is the largest possible value of $s = \cos(\theta)$. Therefore we can state that SLEM equals $\cos(\theta)$ where $\theta$ is the smallest root of (5) in interval $[0, \pi]$. The same result is achievable through (26-a).

**Remark 4.** Recursive equations (25) and (26) are similar to the recursive relations between Chebyshev polynomials of the second kind. This similarity is the original motivation for the choice of $s = \cos(\theta)$. For the choice of optimal weights (29) and (31), equations (25) and (26) are Chebyshev recursions with Chebyshev polynomials defined as $U_i(\cos(\theta)) = a_{m-i}$ for $i = 0, ..., m - 1$.

**Remark 5.** The weight matrix (13) has at least one eigenvalue equal to one (with corresponding eigenvector of all ones). In addition, for all values of $\theta$ satisfying (31), $\cos(\theta)$ is the second largest eigenvalue modulus. Therefore, for the choice of optimal weights, the weight matrix $W$ has one eigenvalue equal to one and all other eigenvalues are less than one in magnitude. According to [12] these are the necessary and sufficient conditions for the convergence of consensus averaging algorithm.

## VI. Quantization

In the realistic scenarios, agents or nodes have to exchange or store quantized data. In an effort to have a more pragmatic analysis, in this section we try to have a comparison between the performance of different weighting methods and quantization schemes in quantized consensus averaging algorithm. The model we have considered for quantized consensus averaging algorithm in this section is

$$x(t+1) = Q(W.x(t)).$$

$x(t)$ is the vector containing states of nodes on the network. $W$ is the weight matrix and by operator $Q$ we mean the quantization scheme. The most simple and common quantization method used in the literature [30] is uniform quantization. In uniform quantization method each value is rounded to the nearest quantization level or equivalently

$$Q(x) = \begin{cases} \lceil x \rceil & if\ x \geq \dfrac{(\lceil x \rceil + \lfloor x \rfloor)}{2}, \\ \lfloor x \rfloor & if\ x < \dfrac{(\lceil x \rceil + \lfloor x \rfloor)}{2}. \end{cases}$$

$\lceil x \rceil$ ($\lfloor x \rfloor$) is the smallest (largest) quantization level greater (smaller) than $x$. Another quantization method used for quantized distributed consensus averaging algorithm is the probabilistic quantization method proposed in [32]. Using probabilistic quantization method $Q(x)$ is a random variable with the following distribution

$$Q(x) = \begin{cases} \lceil x \rceil & \text{with probability } \dfrac{x - \lfloor x \rfloor}{\lceil x \rceil - \lfloor x \rfloor}, \\ \lfloor x \rfloor & \text{otherwise.} \end{cases}$$

In [32] it has been shown that using probabilistic quantization, the nodes in the network eventually achieve consensus. This is due to the fact that using probabilistic quantization the expected value of $Q(x)$ equals $x$.

In Fig. 4 we have illustrated the value of states of nodes in terms of number of iterations for both probabilistic and uniform quantization methods. In the results depicted in Fig. 4 we have considered a symmetric star network with the same topology as Fig. 1(a) with optimal weights and 6 bits for quantization.

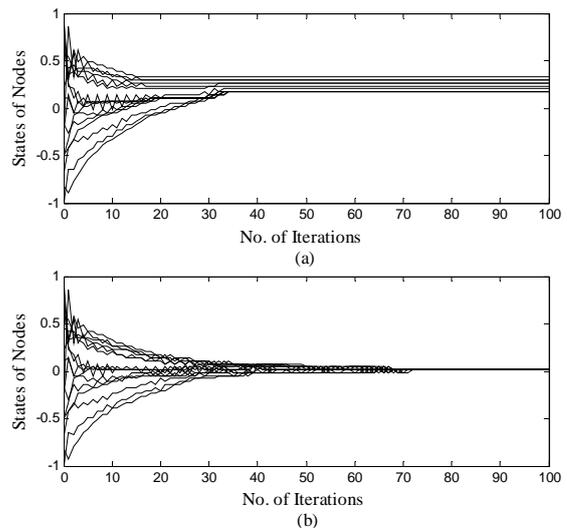

Fig. 4. States of nodes in terms of number of iterations using (a) uniform quantization and (b) probabilistic quantization.

Fig. 4 indicates that with probabilistic quantization, nodes in the network indeed converge to a consensus, as expected.

In order to have a comparison between different weighting methods, in Tables 3, 4 and 5 we present the statistical results for quantized distributed consensus averaging algorithm using probabilistic quantization for different weighting methods and topologies. We have simulated quantized consensus for three network topologies, namely, symmetric and CCS star topologies with $n = 3, m = 2$ and KCS star topology with $n = 3, m = 2, k = 2$. The weighting methods we have considered are maximum degree [12], Metropolis-Hasting [7], best constant [1] weighting methods (Appendix C) and the optimal weights (as provided in section IV). These are commonly used weighting methods in the literature.

The results for uniform quantization are not illustrated in Tables 3, 4 and 5. The main reason is that the number of trials achieving consensus was almost zero due to the use of uniform quantization.

In Tables 3, 4 and 5 we present the statistical results of simulations by four parameters. First parameter ($\psi$) is the percentage of trials where all nodes in the network has reached consensus. Second parameter ($\eta$) is the average number of iterations, it takes for the network to reach consensus. Third ($\mu$) and fourth ($\rho$) parameters are the mean and variance of normalized quantization error. Quantization error is the error between the consensus value and actual mean value of initial states. Normalized quantization error is the quantization error divided by the quantization resolution. $\eta, \mu$ and $\rho$ are only calculated for trials where network has reached consensus. First column of Tables 3, 4 and 5 contains the number of quantization bits. The results depicted in Tables 3, 4 and 5 are generated based on 10000 trials (a different random initial node values is generated for each trial). In our simulations $-1$, $(1)$ is the smallest (largest) quantization level and any value smaller (larger) than $-1, (1)$ is mapped to $-1, (1)$. All other quantization levels are evenly chosen between $-1$ and $1$. We assume that initial states of nodes are distributed uniformly between $-1$ and $1$.

TABLE 3
QUANTIZED CONSENSUS AVERAGING ALGORITHM OVER A SYMMETRIC STAR NETWORK WITH $n = 3, m = 2$.

| No. of Quantization bits | Metropolis | | | | Maximum Degree | | | | Best Constant | | | | Optimal | | | |
|---|---|---|---|---|---|---|---|---|---|---|---|---|---|---|---|---|
| | $\psi$ | $\eta$ | $\mu$ | $\rho$ | $\psi$ | $\eta$ | $\mu$ | $\rho$ | $\psi$ | $\eta$ | $\mu$ | $\rho$ | $\psi$ | $\eta$ | $\mu$ | $\rho$ |
| 4 | 100 | 31.25 | 4.7e-3 | 0.48 | 100 | 27.17 | 1.4e-2 | 0.43 | 100 | 31.18 | 3.1e-3 | 0.594 | 100 | 30.78 | 3.7e-3 | 0.563 |
| 8 | 100 | 56.71 | 1.46e-2 | 1.022 | 100 | 46.98 | 2.54e-3 | 0.86 | 100 | 47.22 | 2.34e-3 | 1.04 | 100 | 46 | 1.21e-2 | 0.9 |
| 16 | 100 | 107.94 | 3.4e-3 | 2.16 | 100 | 87.43 | 9.23e-3 | 1.73 | 100 | 78.98 | 1.2 | 4.7e3 | 100 | 77.72 | 0.6 | 1.57e3 |

TABLE 4
QUANTIZED CONSENSUS AVERAGING ALGORITHM OVER A CCS STAR NETWORK WITH $n = 3, m = 2$.

| No. of Quantization bits | Metropolis | | | | Maximum Degree | | | | Best Constant | | | | Optimal | | | |
|---|---|---|---|---|---|---|---|---|---|---|---|---|---|---|---|---|
| | $\psi$ | $\eta$ | $\mu$ | $\rho$ | $\psi$ | $\eta$ | $\mu$ | $\rho$ | $\psi$ | $\eta$ | $\mu$ | $\rho$ | $\psi$ | $\eta$ | $\mu$ | $\rho$ |
| 4 | 100 | 41 | 4e-3 | 0.464 | 100 | 34.76 | 5.18e-4 | 0.42 | 99 | 55.3 | 5.3e-3 | 0.81 | 99.7 | 42.88 | 2.4e-3 | 0.654 |
| 8 | 100 | 73.86 | 1.14e-3 | 1.02 | 100 | 60.56 | 2.45e-2 | 0.82 | 98.52 | 75.57 | 1.58e-2 | 1.36 | 99.31 | 67.94 | 1.06e-2 | 0.97 |
| 16 | 97.6 | 139.7 | 6.1e-3 | 2.215 | 100 | 112.98 | 1.82e-3 | 1.71 | 94.95 | 112.8 | -0.66 | 1.4e4 | 100 | 104.6 | 0.576 | 2.14e3 |

TABLE 5
QUANTIZED CONSENSUS AVERAGING ALGORITHM OVER A KCS STAR NETWORK WITH $n = 3, m = 2, k = 2$.

| No. of Quantization bits | Metropolis | | | | Maximum Degree | | | | Best Constant | | | | Optimal | | | |
|---|---|---|---|---|---|---|---|---|---|---|---|---|---|---|---|---|
| | $\psi$ | $\eta$ | $\mu$ | $\rho$ | $\psi$ | $\eta$ | $\mu$ | $\rho$ | $\psi$ | $\eta$ | $\mu$ | $\rho$ | $\psi$ | $\eta$ | $\mu$ | $\rho$ |
| 4 | 100 | 24.98 | 4.62e-4 | 0.4 | 100 | 27.2 | 3.93e-3 | 0.402 | 100 | 24.8 | 4.8e-3 | 0.39 | 100 | 24.58 | 4.23e-3 | 0.41 |
| 8 | 100 | 42.17 | 8.25e-4 | 0.72 | 100 | 42.77 | 18e-3 | 0.68 | 100 | 37.96 | 2.67e-3 | 0.66 | 100 | 36 | 13.7e-3 | 0.62 |
| 16 | 100 | 77.04 | 8.27e-3 | 1.4 | 100 | 76.08 | 15.4e-3 | 1.32 | 100 | 64.08 | 2.53e-3 | 1.2 | 100 | 59.41 | 93.9e-3 | 52.7 |

As it is obvious from Tables 3, 4 and 5, probabilistic quantization method assures consensus over the nodes of network. This has been discussed in detail in [32]. Comparing weighting methods it is apparent that optimal weights reach consensus faster than other weighting methods, but not necessarily with the least amount of error. As the number of quantization bits increases the error between consensus value and actual mean value declines, but in the other hand it requires more number of iterations to reach consensus.

## VII. CONCLUSION

In this paper, we have presented the analytical solution for fastest distributed consensus averaging problem over a sensor network with symmetric star topology. This problem has been treated numerically in most of the previous works. We have taken advantage of convexity of FDC averaging algorithm and rich symmetric properties of topologies considered in this paper. The solution procedure consists of stratification method and semidefinite programming. By solving complementary slackness conditions we have reached the characteristic polynomials of weight matrix and by inductive comparison of these polynomials, the optimal weights and SLEM of network are achieved. Also it has been shown that obtained optimal weights for the edges in central part of CCS and symmetric star configurations are independent of the type of their coupled branches. It has been shown by an example that the optimal weight for the edges connected to central node in symmetric star configuration is independent of type of the branches connected to central node. In other words when the path branches in symmetric star topology are replaced by any arbitrary graph, the optimal weights on the edges connected to central node remain unchanged. Same conclusion holds true about the optimal weight for edges of complete core in CCS star topology.

Simulation results confirm better performance of optimal weights compared to other weighting methods (maximum degree, Metropolis and best constant) even under quantization constraints. We are currently extending this method to more generic topologies. Our future directions include the addition of noise and communication delay in asynchronous mode.

## APPENDIX A
### THEOREM (CAUCHY INTERLACING THEOREM) [34]

Let $A$ and $B$ be $n \times n$ and $m \times m$ matrices, where $m \leq n$. $B$ is called a compression of $A$ if there exists an orthogonal projection $P$ onto a subspace of dimension $m$ such that $PAP = B$. The Cauchy interlacing theorem states that if the eigenvalues of $A$ are $\lambda_1(A) \leq \cdots \leq \lambda_n(A)$, and those of $B$ are $\lambda_1(B) \leq \cdots \leq \lambda_m(B)$, then for all $j$,

$$\lambda_j(A) \leq \lambda_j(B) \leq \lambda_{n-m+j}(A).$$

Notice that, when $n - m = 1$, we have

$$\lambda_j(A) \leq \lambda_j(B) \leq \lambda_{j+1}(A).$$

## APPENDIX B
### DEFINITION OF VECTORS $\boldsymbol{\alpha}_i$ & $\boldsymbol{\beta}_i$

For $i = 2, \ldots, m$ the vectors $\boldsymbol{\alpha}_i$ and $\boldsymbol{\beta}_i$ are defined as:

$$\boldsymbol{\alpha}_i(j) = \begin{cases} 1 & \text{for } j = i-1, \\ -1 & \text{for } j = i, \\ 0 & \text{otherwise,} \end{cases} \quad \boldsymbol{\beta}_i(j) = \begin{cases} 1 & \text{for } j = i, \\ -1 & \text{for } j = i+1, \\ 0 & \text{otherwise,} \end{cases}$$

and for $\boldsymbol{\alpha}_1$ and $\boldsymbol{\beta}_1$ we have

$$\boldsymbol{\alpha}_1(j) = \begin{cases} -1 & \text{for } j = 1, \\ 0 & \text{otherwise.} \end{cases} \quad \boldsymbol{\beta}_1(j) = \begin{cases} \sqrt{n} & \text{for } j = 1, \\ -1 & \text{for } j = 2, \\ 0 & \text{otherwise.} \end{cases}$$

## APPENDIX C
### MAXIMUM DEGREE, METROPOLIS-HASTING & BEST

## CONSTANT WEIGHTING METHODS

The Metropolis-Hastings weighting method is defined as:

$$W_{i,j} = \begin{cases} 1/\left(1 + max(d_i, d_j)\right) & j \in N_i, i \neq j \\ 1 - \sum_{j \in N_i} W_{i,j} & i = j \\ 0 & otherwise \end{cases}$$

where $d_i$ and $d_j$ are the degrees of nodes $i$ and $j$, respectively and $N_i$ is the set of immediate neighbors of node $i$.

The Maximum degree weighting method is defined as:

$$W_{i,j} = \begin{cases} 1/\max_k(d_k) & j \in N_i, i \neq j \\ 1 - d_i/\max_k(d_k) & i = j \\ 0 & otherwise. \end{cases}$$

The Best constant weighting method is defined as:

$$W_{i,j} = \begin{cases} \alpha & j \in N_i, i \neq j \\ 1 - d_i \alpha & i = j \\ 0 & otherwise \end{cases}$$

In [20] it has been shown that the optimum choice of $\alpha$ for best constant weighting method is $\alpha^* = 2/(\lambda_1(L) + \lambda_{n-1}(L))$ where $\lambda_i(L)$ denotes the $i$-th largest eigenvalue of $L$ and $L$ is the Laplacian matrix defined as $L = AA^T$ with $A$ as the adjacency matrix of the sensor network's connectivity graph.